\documentclass{article}

\usepackage{graphicx}
\usepackage{xcolor}
\usepackage{amssymb,amsmath}
\usepackage{comment}

\begin{document}

\bibliographystyle{plain}

\title{Multipolynomial Monte Carlo for Trace Estimation in Lattice QCD\footnotemark[1] }

\author{Paul Lashomb\footnotemark[2]
\and Ronald B. Morgan\footnotemark[3] 
\and Travis Whyte\footnotemark[4]
\and Walter Wilcox\footnotemark[5]
}

\maketitle

\renewcommand{\thefootnote}{\fnsymbol{footnote}}
\footnotetext[1]{WW acknowledges support from the Baylor Summer Research Award program.}
\footnotetext[2]{Department of Physics, Baylor University, Waco, TX 76798 ({\tt Paul\_Lashomb@baylor.edu}).}
\footnotetext[3]{Department of Mathematics, Baylor University, Waco, TX 76798 ({\tt Ronald\_Morgan@baylor.edu}).}
\footnotetext[4]{Department of Applied Mathematics and Theoretical Physics, University of Cambridge, Cambridge, CB3 0WA, UK ({\tt tw601@cam.ac.uk}).}
\footnotetext[5]{Department of Physics, Baylor University, Waco, TX 76798 ({\tt Walter\_Wilcox@baylor.edu}).}
\renewcommand{\thefootnote}{\arabic{footnote}}

\begin{abstract}  
Estimating the trace of the inverse of a large matrix is an important problem in lattice quantum chromodynamics.  A multilevel Monte Carlo method is proposed for this problem that uses different degree polynomials for the levels.  The polynomials are developed from the GMRES algorithm for solving linear equations.  To reduce orthogonalization expense, the highest degree polynomial is a composite or double polynomial found with a polynomial preconditioned GMRES iteration.  Added to some of the Monte Carlo pieces is deflation of eigenvalues that reduces the variance.  Deflation is also used for finding a reduced degree deflated polynomial.  The new Multipolynomial Monte Carlo method can significantly improve the trace computation for matrices that have a difficult spectrum due to small eigenvalues. 

\end{abstract}

%\begin{keywords}
% linear equations, eigenvalues, polynomial preconditioning, GMRES, GMRES-DR, deflation, QCD
%\end{keywords}

%\begin{AMS}
%65F15, 15A18
%\end{AMS}

\pagestyle{myheadings}
\thispagestyle{plain}
%\markboth{M. EMBREE, R. B. MORGAN}{Krylov Methods for Rank-one Updated Matrices}

\section{Introduction}

Our goal is to compute $Trace(A^{-1})$ for a large matrix in lattice quantum chromodynamics (QCD) physics.  This is a challenging problem that is generally solved with Monte Carlo sampling.  We give a new method that uses multilevel Monte Carlo with different degree polynomials used to form the levels.  It also incorporates eigenvalue deflation.  We show with some QCD problems that this approach can significantly improve computation compared to simple Monte Carlo.  

$Trace(A^{-1})$ can be computed with a Monte Carlo method using Hutchinson samples $b^{T}A^{-1}b$~\cite{HutchTrace}.  We will use $b$ as a vector of Z4 noise (made of random choices of 1, -1, i and -i), but other choices are possible.  A large system of linear equations $Ax = b$ must be solved for each sampling, and this is the main expense of the method.  Ways of improving this solution have been proposed including deflating eigenvalues~\cite{dF,Qcdconf,StOr,Lan-DR,BiFr} and multigrid methods (see for example~\cite{BaBrBrClMa,FrKaKrLeRo}).  Both of these require significant work for the initial setup.  There is eigenvector computation for deflation, while multigrid finds matrices corresponding to different multigrid levels. 

%\textcolor{red}{There also has been work on speeding up the overall QCD Monte Carlo process.  The approaches mentioned here are related to the new method in this paper.  Multilevel Monte Carlo~\cite{Gi08,Gi15} is applied with the levels being different sized multigrid matrices in~\cite{FrKhRH}.  Probing methods have also been used~\cite{LaStOr,LaSt}.  Deflation of singular values has been proposed for reducing Monte Carlo variance~\cite{QCDsubtr,QCDsubtr2,GaStOr,QCDsubtr3,RoStOr}.  Finally, perhaps the most important previous work for this paper is on using polynomials to reduce the variance of the Monte Carlo~\cite{ThDoLiYe,Walt6,Qcdconf2,QCDsubtr3}. } 

%\textcolor{blue}{
There also has been work on speeding up the overall QCD Monte Carlo process.  The approaches mentioned here are related to the new method in this paper, with the majority of methods developed being single level methods.  Probing methods, which remove the largest elements of the inverse nearest to the diagonal, were proposed in~\cite{LaStOr,LaSt}.  Deflation, which removes the contribution of the largest singular values to the variance, was proposed in~\cite{QCDsubtr,QCDsubtr2,GaStOr,QCDsubtr3,RoStOr}.  
Finally, perhaps the most important previous single level method for this paper is the use of polynomials to reduce the variance of the Monte Carlo~\cite{ThDoLiYe,Walt6,Qcdconf2,QCDsubtr3}.  More recently, multilevel Monte Carlo methods~\cite{Gi08,Gi15} have also been applied to lattice QCD calculations. In~\cite{FrKhRH}, multilevel Monte Carlo is applied with the levels being differences of multigrid matrices projected to the finest grid. Differences of inverses separated by a real shift of the matrix were used in a multilevel fashion in \cite{WhStRoOr, GiHaNaSc}.
%}

In this paper we give a multilevel Monte Carlo method using different degree polynomials of $A$.  We call this Multipolynomial Monte Carlo.  Eigenvalues are used for deflation instead of singular values, and it is shown that the deflation is essential to making this approach effective.  Most important is deflation from the Monte Carlo trace estimation itself.  However, there is also deflation from polynomial preconditioned GMRES for solving linear equations.  Additionally, this paper gives a new deflated polynomial.  So eigenvalue deflation is used in three different ways.  Also needed for efficiency is use of polynomial preconditioned Arnoldi to find the required eigenvalues and eigenvectors.  

Section 2 has quick review of some methods needed in this paper.  Section 3 presents the new multipolynomial method.  Section 4 has discussion of the ways that deflation is used.  The algorithm is given in Section 5 along with testing.  Section 6 finishes up by showing how important deflation and polynomial preconditioning are for efficiency in numerical tests.

\section{Review}

\subsection{Deflation for linear equations}

Deflation in Krylov methods for solving linear equations is a way to remove the effect of small eigenvalues that generally cause slower convergence.  Deflating from the conjugate gradient method was suggested in~\cite{Ni}.  For GMRES deflation, see for example~\cite{GMRES-E,KhYe,ErBuPo,ChSa,Sa95B,BaCaGoRe,BuEr,FrVu,GMRES-DR,PadeStMaJoMa}.  
The GMRES-DR method~\cite{GMRES-DR} computes eigenvectors at the same time it is solving linear equations.  Once the eigenvectors are accurate enough, they speed up the GMRES convergence.  
Examples of QCD papers with deflation are~\cite{dF,Qcdconf,StOr,Lusch,Lan-DR,BiFr}.

GMRES-Proj~\cite{GMRES-DR,gproj} is a deflation method that uses eigenvectors that are already computed, possibly by GMRES-DR.  GMRES-Proj alternates between cycles of GMRES and projections over approximate eigenvectors.  For QCD problems, the left eigenvectors can be found from the right eigenvectors.  Having both left and right gives better deflation, and this is used in~\cite{Qcdconf2, defbi}.  We call the projection using both left and right eigenvectors ``Left-right Projection".  Sometimes it is effective enough that only a single projection is needed.  The algorithm for this is given next.

%\vspace{-.01in}
\begin{center}
\textbf{Left-right Projection}
\end{center}
\vspace{-.18in}
\begin{description}
 \item[0.] Let the current system of linear equations be $A(x-x_0) = r_0$.
 \vspace{-.1in}
 \item[1.] Let $V$ and $U$ be $n$ by $k$ matrices whose columns span the set of approximate right and left eigenvectors, respectively. 
 \vspace{-.1in}
 \item[2.] Form $H = U^* A V$ and $c = U^* r_0$.
 \vspace{-.1in}
 \item[3.] Solve $H d = c$, and let $\hat x = V d$.
 \vspace{-.1in}
 \item[4.] The new approximate solution is $x_p = x_0 + \hat x$, and the new residual is $r = r_0 - A \hat x = r_0 - A V d$.
\end{description}
%\vspace{.15in}

\subsection{Deflation for Trace of $A^{-1}$ Monte Carlo}

Deflation can also be used for Monte Carlo sampling, see~\cite{QCDsubtr,QCDsubtr2,GaStOr,QCDsubtr3,RoStOr}.  The idea is to subtract away components of $A^{-1}$ in order to reduce the size and thus reduce the variance for the Monte Carlo trace computation.  

Assume we have $k$ eigenvalues of $A$ and the associated right and left eigenvectors, so we have $\lambda_i$, $z_i$ and $u_i$ respectively for $i$ from 1 to $k$.  Then the deflated matrix is
\[
A^{-1} - \sum_{i=1}^{k}\frac{1}{\lambda_i} z_i u_i^*,
\]  
However, because $A$ is non-normal, deflating eigencomponents can actually raise the norm and variance.  See Subsection 4.1 for an example of this.  Therefore singular value components were used instead in the references in the previous paragraph.  In~\cite{QCDsubtr2,QCDsubtr3}, singular vectors are found from eigenvectors of $\gamma_5$ times the matrix, and this is referred to as ``Hermitian forced" subtraction.  

%\subsection{Trace of $p(A)$ for low degree polynomials}

\subsection{Polynomial preconditioning using the GMRES polynomial}

Polynomial preconditioning has been suggested to improve Krylov methods for both linear equations and eigenvalues.  See, for instance, 
~\cite{Sa84b,Sa87b,AsMaOt,seed,PPG,LiXiVeYaSa,PPArn,PPGStable}.
For linear equations, polynomial preconditioning replaces $Ax=b$ with 
\begin{align}
Ap(A)y &= b, \nonumber \\  
x &= p(A)y.  \nonumber
\end{align}
The goal is to have $Ap(A)$ with an improved spectrum compared to $A$, so $p(A)$ needs to be an approximation to $A^{-1}$.

In~\cite{seed,PPG}, use of a polynomial from the GMRES algorithm is suggested.  A more stable formulation of this polynomial is given in~\cite{PPArn,PPGStable} using the roots of the GMRES residual polynomial.  Also given is additional stability control featuring added roots for outstanding eigenvalues, however this is not needed for the QCD matrices in this paper.  The algorithms used here for finding the polynomial and for multiplying the polynomial of $A$ with a vector are Algorithms 4 and 3 in~\cite{PPGStable}.

\subsection{Double polynomials}

Polynomial preconditioning with a composite polynomial, called double polynomial preconditioning, is given in~\cite{PPArn,PPGStable}. 
 The first polynomial, called $p_{in}$, is found with GMRES working on matrix $A$, then the second one, $p_{out}$, is generated by polynomial preconditioned GMRES using $p_{in}$.
Once these polynomials have been determined, double polynomial preconditioned GMRES corresponds to these equations: 
\begin{align}
Ap_{in}(A)p_{out}(Ap_{in}(A))z &= b, \nonumber \\  
x &= p_{in}(A)p_{out}(Ap_{in}(A))z.  \nonumber
\end{align}
GMRES is applied to the first equation to find $z$, then the solution to the original system of equations is computed using the second line.
For the purposes in this work, it is important to notice that $p(A) \equiv p_{in}(A)p_{out}(Ap_{in}(A))$ is an approximation to $A^{-1}$.

\subsection{Polynomials for Trace of $A^{-1}$ Monte Carlo}

An idea that has been used to reduce the variance in a QCD Monte Carlo is to find a function $f$ so that $f(A)$ is an approximation to $A^{-1}$ and yet it is possible to compute $Trace(f(A))$.  Then the equation
\[
    Trace(A^{-1}) = Trace(A^{-1} - f(A) ) + Trace(f(A)) %\label{ftr} 
\]
is used with a Monte Carlo of reduced variance for the first trace on the right side of the equation.
This is suggested in~\cite{ThDoLiYe} with Pade approximations to $A^{-1}$.  In~\cite{Walt6}, the approximation is chosen using the geometric series (for matrices this is also called the Neumann series), and this approach is called perturbative subtraction.   A polynomial related to the GMRES polynomial has also been used~\cite{QCDsubtr2,QCDsubtr3}.

\section{Multipolynomial Monte Carlo}

\subsection{Multilevel Monte Carlo with polynomials}

%We want to find $Trace(A^{-1})$ for a large matrix in quantum chromodynamics (QCD) physics.  This trace can be computed with a Monte Carlo method using Hutchinson samples $x^{T}A^{-1}x$~\cite{HutchTrace}, where $x$ is a vector of Z4 noise (made of random choices of 1, -1, i and -i) but other choices are possible.  A large system of linear equations must be solved for each sampling, and this is the main expense of the method.  Ways of improving this solution have been proposed including deflating eigenvalues~\cite{Qcdconf,StOr,Lan-DR,BiFr} (Luscher??) and multigrid methods (see for example~\cite{BaBrBrClMa,FrKaKrLeRo}).  Both of these require significent computation for initial set.  There is eigenvector computation for deflation and finding the different level matrices for multigrid.  

%There also has been work on speeding up the overall QCD Monte Carlo process.  Multilevel Monte Carlo~\cite{Gi08,Gi15} is applied with the levels being different sized multigrid matrices ((refs)).  Probing methods have also been proposed ((refs)).  And as mentioned in Section 2.2, deflation of singular values has been proposed for reducing Monte Carlo variance.  These three approaches just mentioned are all related to the new method in this paper. 
%We give here an approach to improving Hutchison using multilevels with polynomials.  And deflation is implemented with eigenvalues instead of singular valuess.  Also, probing is used for $Trace(p(A))$ with $p$ a low degree polynomial.

Multilevel Monte Carlo~\cite{Gi08,Gi15} uses different levels of approximation to the problem being solved in order to reduce the variance.  It creates several independent Monte Carlo estimations with each except the last one having differences between levels.  The last has a Monte Carlo for the least accurate approximation to the original problem.  Essentially, much of the Monte Carlo sampling is shifted from the original problem to sampling with the approximations that are cheaper to use.
Multilevel Monte Carlo has been applied in QCD~\cite{GiHaNaSc,WhStRoOr}.  Also,~\cite{FrKhRH} has levels being different size multigrid matrices. 

In~\cite{PolyQCD}, it is suggested to use a high-degree polynomial of $A$ to approximate $A^{-1}$ and reduce the variance associated with computing the trace of $A^{-1}$.  Here we turn this into a multilevel Monte Carlo method.  Consider
\begin{equation}
    Trace(A^{-1}) = Trace(A^{-1} - p_1(A) ) + Trace(p_1(A)). \label{onepolytr}
\end{equation}
The two traces on the right can be computed independently with Monte Carlo sampling.  Or if $p_1$ is low degree, the second trace can be computed exactly with probing~\cite{LaStOr,LaSt}.
If $p_1$ is a good approximation to $A^{-1}$, the variance is greatly reduced for the $Tr(A^{-1} - p_1(A))$ part of the Monte Carlo.  Thus very few samples are needed for this part of the trace estimate.  Each sample is expensive, needing solution of linear equations and application of a high degree polynomial, so $p_1$ does need to be high enough degree to be a good approximation to $A^{-1}$.  
High degree polynomials that are generated by the GMRES algorithm are made stable in~\cite{PPArn,PPGStable}, and we will use these polynomials. 
However, \eqref{onepolytr} cannot easily be used with a high degree polynomial, because then finding $Trace(p_1(A))$ requires a Monte Carlo approach that has variance similar to that of the original problem of $Trace(A^{-1})$.  It does gain in that implementing $p_1(A)$ times a vector is generally less expensive than solving the linear equations needed to multiply $A^{-1}$ times a vector.  

To reduce the Monte Carlo cost for $Trace(p_1(A))$, we suggest approximating $p_1$  with a lower degree polynomial, say $p_2$. 
This is done by making $p_2(A)$ another approximation to $A^{-1}$ and thus also an approximation to $p_1(A)$.  The next equation shows how $p_2$ is used so that there are two Monte Carlo's along with the problem of finding the trace of $p_2(A)$:
\[
    Trace(A^{-1}) = Trace(A^{-1} - p_1(A) ) + Trace( p_1(A) - p_2(A) ) + Trace(p_2(A)). %\label{onepolytr}
\]
This is a polynomial version of multilevel Monte Carlo~\cite{He98,Gi08,Gi15}.  In~\cite{HaTr},  multiple Chebyshev polynomials are used in a multilevel Monte Carlo method, however only for symmetric matrices and not for computing the trace of an inverse.  As mentioned above, we instead use GMRES polynomials for our non-Hermitian complex QCD matrices.  Our approach is perhaps easier to set up than a multilevel Monte Carlo method that uses multigrid matrices~\cite{FrKhRH}, since it does not need the development of the matrices.  However, there are also complications for our new approach that will be discussed, such as the need of a high degree $p_1$ and the need of eigenvalue deflation.  

For each polynomial $p_i$, we want
\[
    p_i(A) \approx A^{-1}.  \label{eq:polyappr}
\]
%Of course if $p(A)$ is a low degree polynomial, it cannot be a good approximation, but higher degree polynomials can better approximations. 
The polynomials can be generated from a run of GMRES.  For example, for a degree 100 polynomial $p$, stop GMRES at iteration 101.  We call the GMRES residual polynomial $\pi(\alpha)$.  It is degree 101.  This polynomial can be written as $\pi(\alpha) = 1 - \alpha p(\alpha)$, where $p$ is the polynomial that approximates the inverse of $A$.  In fact, $\pi(0)=1$ and $\pi(\alpha) \approx 0$ over the spectrum of $A$, once GMRES has run far enough.  So $p(\alpha) \approx \frac{1}{\alpha}$ over the spectrum and thus $p(A) \approx A^{-1}$.  

There can be any number of levels in a multilevel Monte Carlo scheme, however there is cost for extra levels because then each individual Monte Carlo sampling must be estimated more accurately.  We will only go as high as using three polynomials which gives four trace calculations.  
Let the polynomials $p_1$, $p_2$ and $p_3$ have degrees be $d_1 > d_2 > d_3.$  The multilevel Monte Carlo with polynomials for the different levels, which we refer to as Multipolynomial Monte Carlo, uses this formula:
\begin{multline}
    Tr(A^{-1}) = Tr(A^{-1} - p_1(A)) + Tr(p_1(A) - p_2(A) ) + \\ Tr(p_2(A) - p_3(A) ) + Tr(p_3(A) ). \label{eq:multpolytr}
\end{multline}
The first three trace computations on the right of equation \eqref{eq:multpolytr} can be performed with Hutchinson Monte Carlo trace estimation.  Here we do the last one exactly with probing.

To find the needed degree of $p_1$, we can monitor the GMRES residual norm while the polynomial is being calculated.  Here we use relative residual norm below $10^{-5}$, which is more than sufficient for an accurate polynomial.
Choice of the other lower degree polynomials is not obvious.  We do not have the information on the variance for the Monte Carlo pieces in \eqref{eq:multpolytr} needed to use optimal choices as suggested in~\cite{Gi15}.  Sampling to estimate the variance with different degree polynomials is suggested in~\cite{HaTr}, but this is too expensive for our problem.  We instead find the degree of $p_2$ with experiments and so do not necessarily have the optimal degree.  For the lowest degree polynomial, $p_3$, we pick degree 4, because the exact trace computation becomes significantly more expensive as the degree increases.  The exact trace can be computed using probing vectors corresponding to a distance-4 coloring of the graph of $A$, or equivalently, a coloring of the graph of $A^4$.

\subsection{Double polynomials}

For larger QCD matrices, the spectrum tends to become more difficult due to more small eigenvalues.  The degree needed for $p_1$ to be a good approximation to $A^{-1}$ becomes large.  Then the GMRES run that finds $p_1$ is expensive due to the orthogonalization of many vectors.  Also, storage of these vectors may not be possible.  So for $p_1$ only, we switch to a double polynomial.  The composite degree needed is generally higher than that needed for a single polynomial, but the orthogonalization expense for finding the double polynomial is much less.

Let $p_{in}$ be a polynomial of degree $dp_{in}$ that is generated by GMRES and approximates $A^{-1}$. Next, polynomial preconditioned GMRES is run using polynomial $p_{in}$.  So PP($dp_{in}+1$)-GMRES is applied with no restarting for $dp_{out}+1$ iterations to find a polynomial $p_{out}$ of degree $dp_{out}$. Then the high degree polynomial $p_1$ is 
\[
p_1(A) = p_{in}(A) p_{out}(Ap_{in}(A)).
\]
The degree of $p_1$ is $dp_1 = (dp_{in}+1)*(dp_{out}+1) - 1.$

As an example of how much orthogonalization expense can be saved with a double polynomial, we look at a matrix of size $n=331{,}776$ from a $12^3$ by 16 configuration.  With GMRES run to relative residual norm less than $10^{-5}$, a $p_1$ polynomial of degree 614 is found. Meanwhile, with PP(30)-GMRES, a polynomial of degree $30*31-1 = 929$ is needed.  But the orthogonalization cost for GMRES run $m$ iterations is about $m^2$ vector operations, so the cost for the polynomial from regular GMRES is about $615^2 = 378{,}225$.  Cost for the two orthogonalizations to find the double polynomial is about $30^2 + 31^2 = 1861$.  A significantly higher degree polynomial is needed for the double polynomial, about $50\%$ higher.  However, the orthogonalization cost is much less, so the double polynomial makes use of high degree polynomials practical for large QCD matrices.

\section{Deflation}

Deflation of eigenvalues is used in three different ways in our method.  There is deflation from Monte Carlos, deflation in the solution of linear equations, and development of a deflated polynomial.  These are described in the following subsections, but first we discuss the computation of the eigenvalues and eigenvectors that are used for the deflation.  We use polynomial preconditioned Arnoldi method (PP-Arnoldi)~\cite{PPArn}.  In some of the tests, we determine the stopping point by monitoring the relative residual of the corresponding polynomial preconditioned GMRES.  

Deflation is more effective if both right and left eigenvectors are used.  For QCD matrices, the left eigenvectors can be computed from the right ones.  Let $(\lambda_i, z_i, u_i)$ be an eigenvalue triplet with eigenvalue $\lambda_i$, right eigenvector $z_i$, and left eigenvector $u_i$ and use the QCD $\gamma_5$ matrix.  If $\lambda_i$ is real, then $u_i = \gamma_5 z_i$.
For an eigenvalue that is part of a complex conjugate pair (unlike for a real matrix, the pair is not exactly conjugate), $u_i = \gamma_5 z_{i+1}$, and $u_{i+1} = \gamma_5 z_i$.

We next give an algorithm for deciding which eigenvalues and eigenvectors to use for deflation.

\begin{center}
\textbf{Determine Eigenvalues and Eigenvectors for Deflation. }
\end{center}
%\medskip
\begin{enumerate}
\vspace{-.22in}
\item {\bf Choose parameters.}  
Pick values for the following:
\vspace{-.05in}

$ndefinit$ = the number of Ritz values that are considered,
\vspace{-.05in}

$imaglim$ = the limit on how small the imaginary part of a Ritz value needs to be in order to be considered a real eigenvalue,
\vspace{-.05in}

$evectol$ = how small the residual norm of both right and left Ritz vectors needs to be for an eigenvalue to be used in deflation.
\vspace{-.05in}
\item {\bf Find possible eigenvalues and eigenvectors.} Compute from the PP-Arnoldi iteration the $ndefinit$ Ritz values and Ritz vectors nearest to the origin.  
\vspace{-.05in}
\item {\bf Identify eigenvalues for deflation.} Beginning with the Ritz value nearest the origin, test if the imaginary part is below $imaglim$ so that the Ritz value will be considered to be approximating a real eigenvalue.  Then if the residual norm is below $evectol$, compute the left eigenvector and if it also has residual norm below $evectol$, accept this as an approximate eigenvalue that will be used for deflation.  For a Ritz value considered to be approximating a complex eigenvalue, test if the next Ritz value is within $imaglim$ of being the conjugate.  If so, consider these two Ritz values to be conjugates and compute both left eigenvectors and test all four left and right for accuracy before accepting.  At the end of the process, set $ndef$ to be the number of Ritz values that are to be used for deflation.
\end{enumerate}

All tests in this paper have $imaglim = 2*10^{-4}$ and $evectol = 10^{-3}$.

\subsection{Deflation for the Monte Carlo}

Multipolynomial Monte Carlo can reduce expense due to moving much of the noise sampling to the cheaper Monte Carlo parts such as $Tr(p_2(A) - p_3(A) )$.  However, the improvement due to having more levels may not be dramatic due to the increased number of samples needed for more Monte Carlos with more demanding error tolerances.  Needed for a big improvement is deflation in the Monte Carlos that have the difference between polynomials.  

We find eigenvalues $\lambda_i$ and corresponding right and left eigenvectors $z_i$ and $u_i$. Then the Monte Carlos for $Tr(p_1(A) - p_2(A) )$ and $Tr(p_2(A) - p_3(A) )$ have deflation applied.  For the first of these, this is
\[Tr \Bigl(p_1(A) - p_2(A) - \sum(p_1(\lambda_i) - p_2(\lambda_i)) z_i u_i^*\Bigr).
\]  
The deflation reduces the variance for Monte Carlo sampling.  We correct for the deflation part by finding its trace exactly using that the trace of $z_i u_i^*$ is the inner product between the vectors.  

Note that when simply doing a Monte Carlo for $Trace(A^{-1})$, deflating eigenvalues is not very effective because of non-normality effects.  For a non-normal matrix, there can even be an increase in the norm of the inverse as some spectral components are removed.  The norm of $A^{-1}$ will not go up if singular value decomposition components are instead removed, hence they have been previously used~\cite{QCDsubtr,QCDsubtr2,GaStOr,QCDsubtr3,RoStOr}.  Nevertheless, we remove eigenvalue components, because the singular values and vectors of a polynomial of $A$ are not necessarily related to those of $A$.  Also, eigenvalues work well in our context because the polynomials all approximate $A^{-1}$ but differ mostly at the small eigenvalues that we are deflating.  So removing these components significantly reduces the variance.  
We now study this eigenvalue deflation further with a random matrix.  Some of the same effects as for QCD matrices happen with random matrices. 

{\it Example 1.} Let $B$ be a size $n = 500$ matrix whose entries are random Normal(0,1) and let $A = B + 21*I$.  Figure 1 shows the eigenvalues of $A$ on the left.  Then the right side has a closeup of the eigenvalues nearest the origin, numbered in order of distance from the origin.  
We look at removing eigencomponents, one at a time.  The green numbers (1, 2, 4, 5, 6, 7 and 24) on the right side of the figure indicate that the norm of the deflated matrix goes significantly down with the removal of that component.  The red ones (3 and 19) show where it goes significantly up.  
Figure 2 shows, on the top, the norm of the matrix $A^{-1}$ with accumulating eigenvalue deflation using the right and left eigenvectors.  So at eigenvalue $k$ on the figure, the quantity $\| A^{-1} - \sum_{i=1}^k \frac{1}{\lambda_i} z_i u_i^*\|$ is plotted.
The bottom of the figure has inner products of the right eigenvectors with the right eigenvectors $z_3$ and $z_{19}$.  This shows that $z_3$ is related to its neighboring eigenvectors, especially $z_1$, $z_2$, $z_6$, and $z_7$.  
We discuss two interesting features of the curve in the top plot.  It comes down at eigenvalue 7, which finishes a group of the smallest 7 eigenvalues that somewhat stand out from the others.  Second, the norm goes way up when the 19th is deflated and then down when the 24th is deflated.  These two are very close together, see Figure 1, and are coupled together as is shown by their inner product in the bottom part of Figure 2.  The inner product between their eigenvectors is very near 1.  If two or more eigenvalues are coupled together, they all need to be deflated before the norm will drop.  By contrast, deflating the 20 smallest singular value components causes a larger drop in the norm, down to 0.31 compared to 3.85 with deflating 20 eigencomponents.  

\begin{figure}
\vspace{-2.6in}
\hspace{-.9in}
\includegraphics[scale=.75]{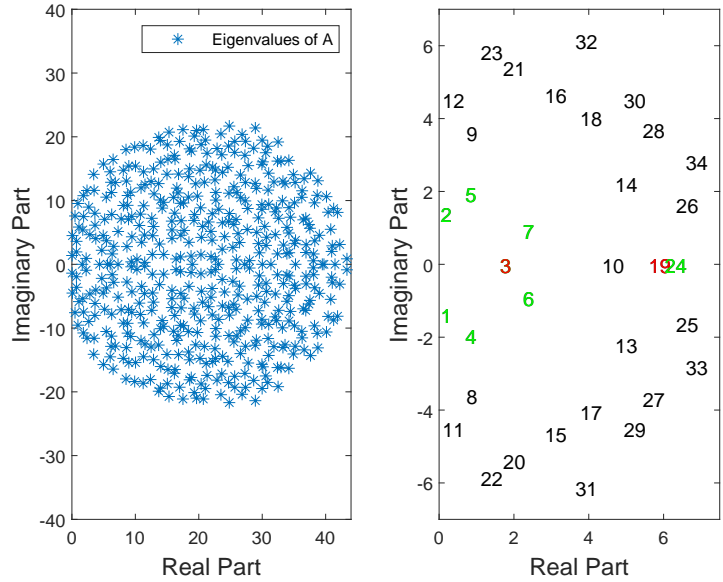}
\vspace{-2.8in}
\caption{Matrix of size $n = 500$ is generated random normal and then shifted.  Eigenvalues are shown on the left, and the right side has a close-up of the small eigenvalues numbered in order of magnitude.}
\label{fig:Ran1}
\end{figure}

\begin{figure}
\vspace{-2.6in}
\hspace{-.9in}
\includegraphics[scale=.75]{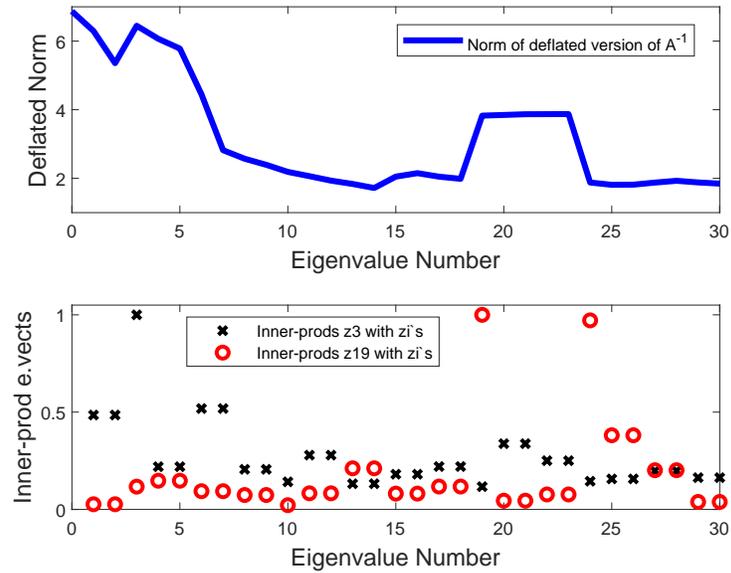}
\vspace{-2.8in}
\caption{Random matrix with $n = 500$.  The top part has the norm of $A^{-1}$ after eigencomponents are subtracted.  The bottom part has inner-products between the right eigenvector $z_3$ and the first 30 right eigenvectors and similarly for $z_{19}$. }
\label{fig:Ran2}
\end{figure}

We now consider the matrix $p_{200}(A)-p_{100}(A)$, where $p_{200}$ is a degree 200 polynomial approximating $A^{-1}$ found with GMRES, and $p_{100}$ is similar with degree 100.
Figure 3 shows on top the eigenvalues of the matrix $p_{200}(A)-p_{100}(A)$.  The largest eigenvalues correspond to small eigenvalues of $A$. Five eigenvalues are well-separated from the rest of the spectrum and the norm of the deflated matrix comes well down with one deflated and also significantly down when all five deflated.  For this difference between polynomials, effective deflation only requires a few eigenvalues.  Also, the deflation does not seem as susceptible to causing an increase in norm as when deflating from $A^{-1}$.  

\begin{figure}
\vspace{-2.6in}
\hspace{-.9in}
\includegraphics[scale=.75]{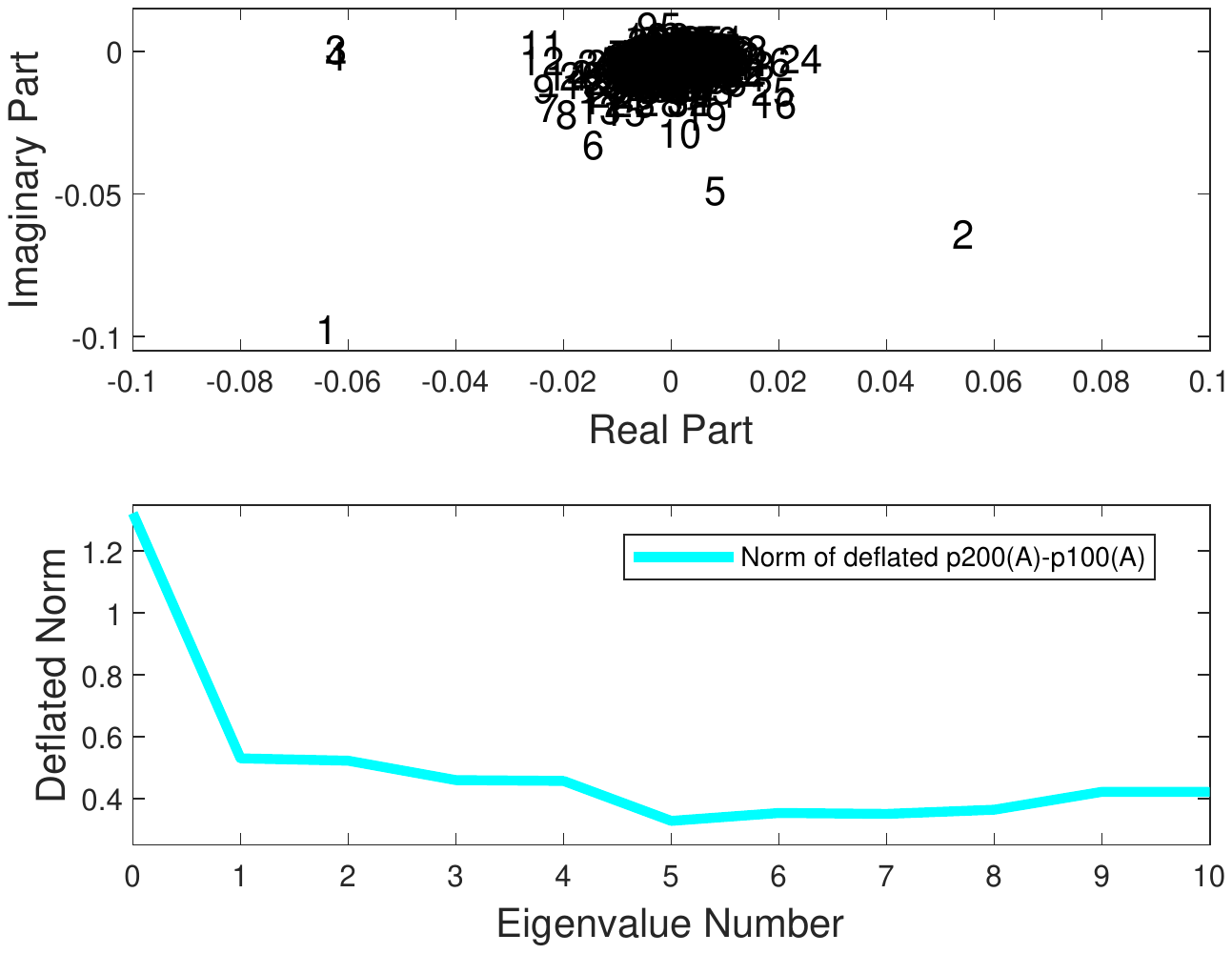}
\vspace{-2.8in}
\caption{Random matrix with $n = 500$.  The top has the eigenvalues of the matrix $p_{200}(A)-p_{100}(A)$, the difference between polynomials of $A$ of degree 200 and 100.  The lower part has the norm of the difference matrix with increasingly more of the large eigencomponents removed.}
\label{fig:Ran3}
\end{figure}

\subsection{Deflation in the solution of linear equations}

The second way deflation is used is in the solution of linear equations.  Left-Right Projections (see Subsection 2.1) are used for the deflation of polynomial preconditioned GMRES.  The polynomial is the same one that is used in the initial PP-Arnoldi iteration that computes the eigenvalues.  Possibly only one projection is needed, as the PP-GMRES may not need restarting due to the deflation and the polynomial preconditioning.

\subsection{Deflated polynomial}

Like the first type of deflation, this third type is a new development.  We generate $p_1$ as a deflated polynomial.  This high degree double polynomial can have significantly reduced degree with the deflation.  

To find a deflated polynomial, first a single Left-Right Projection is done before PP-GMRES is solved to a specified residual tolerance.  The resulting polynomial is then calculated.  For applying this polynomial to vectors, there must first be a projection.

\section{Algorithm and Testing}

We next give the algorithm for our multipolynomial approach.  

\begin{center}
\textbf{Multipolynomial Monte Carlo Trace Estimation \\ with GMRES polynomials. }
\end{center}
%\medskip
\begin{enumerate}
\item {\bf Setup} Choose these parameters:
\begin{enumerate}
\item $\epsilon$ = level of standard error needed by the Monte Carlo,
\vspace{-.05in}
\item $dp_{in}$ = degree of $p_{in}$, $dp_2$ = degree of $p_2$, and $dp_3$ = degree of $p_3$,
\vspace{-.05in}
\item $rtolev$ = relative GMRES residual tolerance used to determine when to stop PP-Arnoldi for finding eigenvalues,  
\vspace{-.05in}
\item $rtolpoly1$ = relative residual tolerance for finding the degree of the deflated double polynomial, so for finding $p_{out}$ with PP-GMRES,
\vspace{-.05in}
\item $rtolle$ = relative residual tolerance for solving linear equations,
\vspace{-.05in}
\item $nevch$ = number of small Ritz values that are checked.
\vspace{-.05in}
\item $m$ = restart parameter for GMRES.
\end{enumerate}

\item {\bf Find polynomials and eigenvectors}
\begin{enumerate}
\item Let $maxit= max\{dp_{in}+1,dp_2+1, dp_3+1\}$.  Run one cycle of \\ GMRES($maxit$) with a random normal starting vector of norm one.  Along the way, determine the polynomials $p_3$, $p_2$, and $p_{in}$ by computing harmonic Ritz values at the proper points. 
\item Run PP($dp_{in}+1$)-GMRES/Arnoldi to relative GMRES residual tolerance of $rtolev$ with the same starting vector as the previous step.  Compute the $nevch$ smallest Ritz pairs (regular Ritz, not harmonic), find left eigenvectors and select ones that have converged enough (see algorithm in Section 4).  Let $ndefl$ be the number of accepted eigenvalues.
\item Find deflated polynomial.  
\begin{enumerate}
\item With a Z4 noise starting vector, deflate eigenvalues with a Left-right Projection using the accepted right and left eigenvectors.
\item Apply PP($dp_{in}+1$)-GMRES until it achieves relative residual tolerance of $rtolpoly1$.
\item Find the polynomial by computing harmonic Ritz values.  
\end{enumerate}
\end{enumerate}
\medskip
\item {\bf Multilevel Monte Carlo} 
\begin{enumerate}
\item Set tolerance for standard error in the first Monte Carlo to $\frac{1}{\sqrt{3}} \epsilon$.  Apply Monte Carlo to compute $Tr(A^{-1}-p_1(A))$, using Z4 noise vectors.  Absolute values are taken of all computed quantities $b^*(A^{-1}-p_1(A))b$.  To solve the linear equations for $A^{-1}b$, use deflated PP($dp_{in}$ + 1)-GMRES(m).  Deflation is with a Left-right Projection using the accepted right and left eigenvectors before every cycle (there may be only one cycle).    
Note that the $p_1(A)$ applied to the first Z4 vector has already been done in finding $p_1$, saving expense here.
We check for acceptable standard error starting at two noises.  Let $currerr$ be the current standard error at the end of the Monte Carlo.   
\item Set tolerance for standard error in the second Monte Carlo to \\ $\frac{1}{\sqrt{2}} \sqrt{\epsilon^2-currerr^2}$.  Apply Monte Carlo to compute $Tr(p_1(A)-p_2(A))$, using Z4 noise vectors.  We check for acceptable standard error starting at three noises and check every three noises thereafter.  Compute $currerr$, the current standard error, by square root of sum of the squares of the standard errors for the two Monte Carlo's.   
\item Set tolerance for standard error in the second Monte Carlo to \\ $\sqrt{\epsilon^2-currerr^2}$.  Apply Monte Carlo to compute $Tr(p_2(A)-p_3(A))$, using Z4 noise vectors.  We check for acceptable standard error starting at six noises and check every three noises thereafter.  Compute $currerr$, the current standard error, by square root of the sum of the squares of the standard errors for the three Monte Carlo's.  
\item Find $Tr(p_3(A))$ exactly using probing.
\item The standard error for the entire multipolynomial process is $currerr$.
\end{enumerate}

\end{enumerate}
\vspace{.08in}

We generally do not want to restart the polynomial preconditioned GMRES in the algorithm, because that can slow convergence.  So it is beneficial to pick the degree of the $p_{in}$ polynomial high enough to avoid this.  

{\it Example 2.} All experiments in this paper are run on a desktop Dell Optiplex computer.  We test with a QCD lattice of size $12^3$ by $16$ that gives an associated quenched Wilson-Dirac fermion matrix of size $n=331{,}776$.  Parameters used are $\beta=6.0$ and $\kappa$ at 0.157 which is essentially $\kappa$-critical (zero quark mass).  We compare several different approaches.  We choose a tolerance of $\epsilon = .0005*V$, where V is the volume of the lattice, which provides a constant tolerance for the variance of the trace per lattice site and is consistent across lattices of different volumes.  So here $\epsilon = 0.0005 * 12^3 * 16$.  Let $rtolev = 10^{-12}$, $rtolpoly1 = 10^{-5}$ and $rtolle = 10^{-6}$.  Tables 1 through 3 test with portions implemented of the Multipolynomial Trace Algorithm.
 
Table 1 has a test of the basic multipolynomial algorithm with no deflation or polynomial preconditioning.  This is compared to regular Hutchison Monte Carlo that uses restarted GMRES(50) to solve the linear equations.  The multipolynomial method with three polynomials uses polynomials of degree 929, 300 and 4.  The highest degree polynomial is a double polynomial with inner polynomial $\phi_{in}$ of degree 30.  The results show that Multipolynomial Monte Carlo can reduce costs.  Hutchinson takes 3.49 days. As polynomials are added, the time comes down and goes as low as 7.35 hours.   The number of noise vectors is 888 for $Tr(A^{-1})$.  This is reduced to 694 for $Tr(A^{-1} - p(A))$ with $p$ a polynomial of degree 4, and then $Tr(p(A))$ is computed exactly with probing.  However, with two or three polynomials, the total number of noise vectors needed goes up.  This is expected for Multilevel Monte Carlo, because the several Monte Carlo's must be solved to stricter tolerance.  The cost savings comes from many of the noises being used with low degree polynomials that are cheaper than $A^{-1}$.

Table 2 has results with more efficient solution of the linear equations.  PP(30)-Arnoldi is used to compute 13 small eigenvalues (run until the GMRES relative residual goes below $rtolev = 10^{-12}$).  The polynomial degrees with three polynomials are 359, 100 and 4.  Linear equations are solved with deflated PP(30)-GMRES(50).  This reduces the Hutchinson time from three and a half days to three and a half hours.  Hutchinson is then efficient enough that the multipolynomial method does not reduce the time, even with a deflated polynomial being used for $p_1$.

Table 3 adds deflation to the Monte Carlo pieces $Tr(p_1(A) - p_2(A))$ and $Tr(p_2(A) - p_3(A))$, again with the 13 eigenvalues.  This greatly reduces Monte Carlo variance and the number of noise vectors needed.  The time comes down to 20 minutes for the multipolynomial method with three polynomials.  Deflation makes a difference in solving the linear equations and for the deflated polynomial, but it is especially vital for the Monte Carlo samplings.

\begin{table}
\caption{ Trace for $12^3 \times 16$ lattice with error tolerance $0.0005*12^3*16$.  GMRES(50) is used to solve linear equations.  The degree $30*31-1 = 929$  polynomial is a double polynomial. }
\begin{center}
\begin{tabular}{|c|c|c|c|c|c|}  \hline\hline
Method                          & Noise vectors     & Time          & MVP's             \\ \hline \hline
Hutchinson, GMRES(50)           & 888               & 3.49 days     & $2.52*10^{6}$     \\ \hline
1 Poly, $deg=4$                 & 694               & 2.71 days     & $1.99*10^6$       \\ \hline
2 Poly's, $deg = 929, 4$        & 2+1572= 1574      & 11.5 hours    & $1.49*10^6$       \\ \hline
3 Poly's, $deg = 929, 300, 4$   & 2+138+2367=2507   & 7.35 hours    & $8.90*10^5$       \\ \hline \hline
\end{tabular}
\end{center}
\label{Tab:Tr1}
\end{table}
\vspace{-.1in}

\begin{table}
\caption{ Trace for $12^3 \times 16$ lattice with error tolerance $0.0005*12^3*16$.  Deflated PP(30)-GMRES(50) used to solve linear equations.  A deflated double polynomial of degree $30*12-1 = 359$ is used for the large polynomial.  For both types of deflation, 13 eigenvalues are used along with their right and left eigenvectors.}
\begin{center}
\begin{tabular}{|c|c|c|c|c|c|}  \hline\hline
Method                          & Noise vectors     & Time          & MVP's             \\ \hline \hline
Hutch., defl. PP(30)-G(50)      & 888               & 3.43 hours    & $4.20*10^5$       \\ \hline
1 Poly, $deg=4$                 & 694               & 2.71 hours    & $3.31*10^5$      \\ \hline
2 Poly's, $deg = 359, 4$        & 2+1572=1574       & 4.50 hours    & $5.81*10^5$       \\ \hline
3 Poly's, $deg = 359, 100, 4$   & 2+1107+1815=2924  & 5.50 hours    & $7.01*10^5$       \\ \hline \hline
\end{tabular}
\end{center}
\label{Tab:Tr2}
\end{table}
\vspace{-.1in}

\begin{table}
\caption{ Trace for $12^3 \times 16$ lattice with error tolerance $0.0005*12^3*16$.  Now with deflation also in the Monte Carlo traces between two polynomials.}
\begin{center}
\begin{tabular}{|c|c|c|c|c|c|}  \hline\hline
Method                          & Noise vectors     & Time          & MVP's             \\ \hline \hline
Hutch., defl. PP(30)-G(50)      & 888               & 3.43 hours    & $4.20*10^5$       \\ \hline
%((not sure:)) 1 Poly, $deg = 4$               & ??                & ?? hours      & $??*10^5$       \\ \hline
2 Poly's, $deg = 359, 4$        & 2+186=188         & 36.1 min's    & $7.67*10^4$       \\ \hline
3 Poly's, $deg = 359, 100, 4$   & 2+6+276=284       & 20.0 min's    & $3.47*10^4$       \\ \hline \hline
\end{tabular}
\end{center}
\label{Tab:Tr3}
\end{table}
\vspace{.2in}

We finish this example by considering the multilevel Monte Carlo cost formula~\cite{Gi15}, given here for three levels,
\begin{equation}
C = \frac{1}{\epsilon^2} \big( \sum_{l=0}^{2} \sqrt{V_l C_l} \big)^2,     \label{eq:mlevcost}
\end{equation}
where $\epsilon$ is the desired accuracy, and $V_l$ and $C_l$ are variance and cost at different levels.  Cost for $Tr(p_3(A))$ is not considered here.  This multilevel cost formula is compared to Hutchinson Monte Carlo cost 
\[C = \frac{1}{\epsilon^2} V_0 C_0 .\]
Deflation and polynomial preconditioning is used for all the tests.  We approximate the cost for each level by using the number of matrix-vector products and do not include the setup cost of generating polynomials and eigenvalues.  Both cost and variance for the different levels are estimated at each level by averaging the results with 50 noise vectors.  Here $\epsilon  = 0.0005 * 12^3 * 16.$  Table 4 has the pieces of the cost calculations and also the final cost $C$.  For Hutchinson, $C = 4.25*10^5$ and for the multilevel method, the overall cost in \eqref{eq:mlevcost} is $2.71*10^4$.  This shows that the multipolynomial method is expected to be an improvement, and roughly matches with the improvement in Time and MVP's in Table 3.  

\begin{table}
\caption{ Multilevel cost calculation for $12^3 \times 16$ lattice with error tolerance $0.0005*12^3*16$. }
\begin{center}
\begin{tabular}{|c|c|c|c|c|c|}  \hline\hline
Method                  & Level         & Variance          & Cost per noise    & Overall Cost  \\ \hline \hline
Hutch. (no poly)        & 0             & $1.81*10^5$       & 450               & $4.25*10^5$   \\ \hline \hline

Multipoly               & 0             & $1.37*10^{-3}$    & 825               &               \\ \hline
                        & 1             & $9.56*10^1$       & 460               &               \\ \hline
                        & 2             & $4.11*10^4$       & 104               & $2.71*10^4$   \\ \hline \hline
\end{tabular}
\end{center}
\label{Tab:Tr3}
\end{table}
%\vspace{.2in}

{\it Example 3.}  We test with 10 larger matrices from $24^4$ configurations.  The matrices are size $n=3{,}981{,}312$.  The desired error is set at $\epsilon = 0.0005 * 24^4$.
We use two versions of the multpolynomial approach.  Both are with three polynomials.  The first polynomial is from deflated PP(50)-GMRES solved to residual norm below $10^{-5}$.  This gives $p_1(A)$ that is a very accurate approximation to $A^{-1}$, so the Monte Carlo for $Tr(A^{-1} - p_1(A))$ requires only two noises and achieves far more than the requested accuracy.  The second and third polynomials are degree 200 and degree 4.  The standard error for the second Monte Carlo is checked after three noises and then rechecked after every three.  The third Monte Carlo is checked after six noises and then every third thereafter.  
The first tests in columns 2 through 4 of Table 5 compute the eigenvectors by running PP(50)-Arnoldi until the corresponding GMRES relative residual norm goes below $10^{-12}$.  From the subspace thus generated, 30 right eigenvectors are computed corresponding to the smallest Ritz vectors, the left eigenvectors are found and the eigenvalue is accepted if residual norms are below $10^{-3}$ (see algorithm in Section 4).  Ritz values are identified as real if the imaginary parts are below $2*10^{-4}$.  An average of 22.4 small eigenvalues are used.  The time and matrix-vector products for computing $Tr(p_3(A))$ with probing are included in the table (the time averages 0.45 hours).
The time for the whole multipolynomial process varies greatly from 3.9 hours up to 16 hours.  This is because the Monte Carlo deflation is sometimes more effective than for other cases.  The number of noise vectors needed for the second Monte Carlo varies from 3 to 66, and every noise vector is expensive because it requires multiplication by the high-degree polynomial $p_1$ (and by $p_2$).    

\begin{table}
\caption{ Trace for $24^4$ lattices with error tolerance $0.0005*24^4$.  Two different amounts of eigenvector calculation are used.  Time is given in hours. }
\begin{center}
\begin{tabular}{|c|c|c|c||c|c|c|c|}  \hline\hline
& \multicolumn{3}{|c||}{Eigenvalues found from} &  \multicolumn{3}{|c|}{Extra deflation, eigenval's from }  \\ 
 & \multicolumn{3}{|c||}{PP(50)-Arnoldi to $rn < 10^{-12}$} &  \multicolumn{3}{|c|}{PP(70)-Arnoldi for 150 it's }  \\ \hline
 Matrix & Noises, 3     & Time,     & MVP's         & Noises, 3  & Time,    & MVP's         \\ 
        & M. Carlo's    & Hours     &               & M. Carlo's & Hours    &              \\ \hline \hline
1       & 2, 3, 60      & 3.93      & $3.72*10^4$   & 2, 3, 15  & 2.85      & $2.59*10^4$   \\ \hline
2       & 2, 3, 105     & 4.93      & $4.52*10^4$   & 2, 3, 12  & 2.81      & $2.56*10^4$   \\ \hline
3       & 2, 3, 75      & 3.98      & $3.82*10^4$   & 2, 3, 42  & 3.46      & $3.18*10^4$   \\ \hline
4       & 2, 6, 60      & 4.23      & $4.04*10^4$   & 2, 3, 24  & 3.05      & $2.81*10^4$   \\ \hline
5       & 2, 66, 177    & 16.3      & $1.68*10^5$   & 2, 3, 30  & 3.22      & $2.88*10^4$   \\ \hline
6       & 2, 6, 42      & 3.85      & $3.57*10^4$   & 2, 3, 21  & 2.96      & $2.68*10^4$   \\ \hline
7       & 2, 39, 132    & 11.0      & $1.04*10^5$   & 2, 3, 21  & 2.93      & $2.70*10^4$   \\ \hline
8       & 2, 3, 63      & 3.93      & $3.62*10^4$   & 2, 3, 27  & 3.13      & $2.82*10^4$   \\ \hline
9       & 2, 39, 195    & 11.8      & $1.20*10^5$   & 2, 3, 27  & 3.15      & $2.87*10^4$   \\ \hline
10      & 2, 3, 108     & 4.46      & $4.35*10^4$   & 2, 3, 30  & 3.10      & $2.93*10^4$   \\ \hline
\hline
\end{tabular}
\end{center}
\label{Tab:Tr3}
\end{table}
%\vspace{.2in}

Next, we check whether it is worth spending more effort developing the eigenvalues for deflation.  We use a degree 69 polynomial for $p_{in}$, and run PP(70)-Arnoldi for 150 iterations (stopping at that set point instead of at a GMRES residual level).  This takes a total of 10{,}500 matrix-vector products which compares to an average of 6250 matrix-vector products used just above when going to $10^{-12}$ (this varies considerably for the 10 matrices, between 4850 and 7000).  Next, the 60 smallest Ritz values and vectors are tested for accuracy.  An average of 52.9 approximate eigenvalues are accepted.  Most of them are very accurate with residual norms around $10^{-15}$.  This makes the decisions more clear-cut for determining which eigenvalues to use for deflation.  The last three columns of Table 5 have the results deflating these eigenvalues; we call it ``extra deflation".  The overall time goes down for all tests even though more time is spent on the initial steps of finding eigenvalues and polynomials before the actual Monte Carlos.

Table 6 has a breakdown of the times for six stages of the multipolynomial process for matrix 8.  With extra deflation, more time is used for finding the eigenvalues and eigenvectors, but much less is needed for the last Monte Carlo.  The first two Monte Carlos also are quicker because $p_1$ is lower degree (see Subsection 6.1).  Also, the linear equations are solved quicker.

\begin{table}
\caption{ Breakdown of times for matrix 8.}
\begin{center}
\begin{tabular}{|c|c|c|c|c|c|}  \hline\hline
Part of multipoly method            & Deflation with $10^{-12}$     & Extra Deflation       \\ 
                                    & Time (minutes)                & Time (minutes)        \\ \hline \hline
Find single polynomials             & 16.6                  & 16.6          \\ \hline
Find eigenvalues and vectors        & 44.1                  & 73.1          \\ \hline
Find deflated double polynomial     & 8.5                   & 4.6           \\ \hline
1st Monte Carlo                     & 30.8                  & 15.5          \\ \hline 
2nd Monte Carlo                     & 17.8                  & 16.0          \\ \hline
3rd Monte Carlo                     & 81.0                  & 35.0          \\ \hline 
Trace of $p_3$                      & 26.7                  & 27.0          \\ \hline \hline
\end{tabular}
\end{center}
\label{Tab:Tr1}
\end{table}

The results in Table 5 are much more consistent with extra deflation.  Looking at the deflated eigenvalues helps explain this.  We examine eigenvalues computed for two of the matrices, 1 and 5, both without and with the extra deflation.  Figure 4 has a plot of the accepted eigenvalues with (red) asterisks computed by PP(50)-Arnoldi solved to relative GMRES residual tolerance of $10^{-12}$.  Then the (black) circles are the eigenvalues with extra deflation, so from PP(70)-Arnoldi run for 150 iterations.  
This figure shows us a couple things, first explaining why Hutchinson method is much slower for Matrix 1 than 5 (338 noises vs. 76 noises). Matrix 1 has much smaller eigenvalues.  Second, it shows why for Matrix 5, the new multipolynomial method is not effective with regular deflation.  Removing the red asterisk eigenvalues leaves other eigenvalues that are almost as close to the origin.  But then with extra deflation, all the black circles are also removed which takes care of all quite small eigenvalues.  With that, the multipolynomial method takes only 4.5 hours.  
The multipolynomial method with regular deflation takes longer than Hutchinson (it is 16 hours for new method and 8.5 hours for Hutchinson with extra deflated PP(70)-GMRES).  However, this is possible, because of the need for each individual Monte Carlo in a multilevel Monte Carlo to be solved to a further tolerance than a single Monte Carlo would need to be.  In this case, a total of 245 noises are needed for the new method with regular deflation, compared to 76 noises for Hutchinson.  With extra deflation, only 41 noises are needed.  This shows how important deflation is to the two Monte Carlos steps involving a difference of polynomials.
 
\begin{figure}
\vspace{-2.6in}
\hspace{-.9in}
\includegraphics[scale=.75]{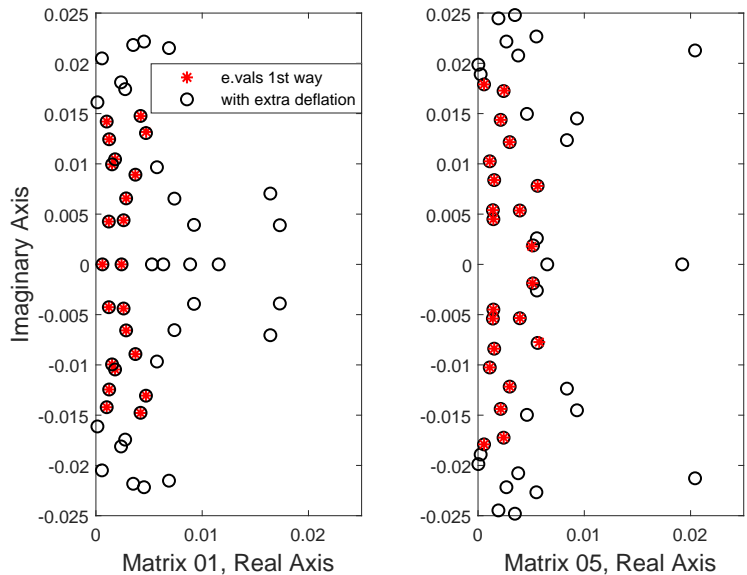}
\vspace{-2.8in}
\caption{Deflated eigenvalues for $24^4$ matrices 1 and 5.  Asterisks are eigenvalues for regular deflation ($rtolev = 10^{-12}$) and circles are for extra deflation eigenvalues.}
\label{fig: Spec244}
\end{figure}
 
We finish off this example with a comparison of Multipolynomial Monte Carlo with Hutchison.  Linear equations are solved for Hutchison  using PP(70)-GMRES with extra deflation.  Table 7 has the results with averaging over all 10 matrices.  The time is reduced by a factor of more than six with the extra deflation version of the multipolynomial method.  We note that the Hutchison method takes an average of 19.6 hours, but this would take much longer if polynomial preconditioning and deflation were not used.  This is discussed in the next section.  

\begin{table}
\caption{ Trace for $24^4$ lattices with error tolerance $0.0005*24^4$. Compare Hutchinson with Multipolynomial Method.  Average with matrices from 10 configurations.}
\begin{center}
\begin{tabular}{|c|c|c|c|c|c|}  \hline\hline
Method                              & Noise vectors     & Time          & MVP's             \\ \hline \hline
Hutch., defl. PP(70)-GMRES          & 188               & 19.6 hours    & $2.02*10^5$       \\ \hline
3 Poly's, $10^{-12}$ for deflation  & 2, 17.1, 101      & 6.84 hours    & $6.68*10^4$       \\ \hline
3 Poly's, extra deflation           & 2, 3, 24.9        & 3.07 hours    & $2.80*10^4$       \\ \hline \hline
\end{tabular}
\end{center}
\label{Tab:Tr3}
\end{table}
\vspace{.2in}

\section{Effectiveness of the Deflated Polynomial and the Modified GMRES and Arnoldi Methods}

The results given in Table 7 for both the multipolynomial method and regular Hutchinson depend on some of the details of the methods used.  For instance, it was mentioned earlier how much the double polynomial reduces orthogonalization costs.  In this section, we highlight three other parts of the methods and show how they effect the computational time. 

\subsection{Deflated polynomial}

The Multipolynomial Monte Carlo method needs a high-degree polynomial $p_1$ in order for $p_1(A)$ to be a very good approximation to $A^{-1}$.  However, the use of a deflated polynomial makes the degree much lower and thus the expense of using the polynomial much less.

{\it Example 3, continued.} To show the importance of a deflated polynomial for $p_1$, we give data from the tests in Example 3 on the degree of $p_1$ with deflation compared to without.  If deflation was not used in developing the polynomial, the degree of the double polynomial $p_1$ would be between $33*50-1 = 1649$ and $79*50-1 = 3949$, with an average of $59*50-1 = 2949$.  With the first amount of deflation (left half of Table 5), $p_1$ has much lower degree, between $24*50-1 = 1199$ and $35*50-1 = 1749$ with an average of $27.2*50-1 = 1360$.  Finally, the extra deflation polynomials are even lower with degree either $10*70-1 = 699$ or $11*70-1 = 769$ and an average of $10.4*70 = 728$.  The deflated polynomial significantly changes the cost of using the $p_1$ polynomial to multiply vectors, by a factor of about four with extra deflation. This reduces the time for both of the first two Monte Carlos steps in the multipolynomial method.

\subsection{Polynomial preconditioning and deflation for solving linear equations}

The solution of linear equations requires the majority of time for the Hutchison method.  In the multipolynomial method, a couple of systems need to be solved.  We look now at how much this is improved by the combination of polynomial preconditioning and deflation.  Matrices 1, 5 and 6 from Example 3 are used as a representative sample.  The first method is GMRES(50), so GMRES with restarting every 50 iterations.  Then the second method has polynomial preconditioning added, specifically PP(50)-GMRES(50).  The third approach is GMRES(50) with extra deflation (``extra" as in Example 3), so a Left-right Projection is applied between GMRES cycles.  Next, PP(50)-GMRES(50) is combined with regular deflation and then PP(70)-GMRES(50) has extra deflation for the fifth method.  Table 8 has results from the average of solving two systems with right-hand sides from Z4 noise to relative residual norms below $10^{-6}$.  Both polynomial preconditioning and deflation significantly reduce the time.  Finding the eigenvectors for deflation is not included in the timings.  The first two methods do not need this computation, so PP(50)-GMRES(50) would be the best approach if one was only solving a couple of systems of linear equations.  However, since deflation is used in other parts of Multipolynomial Monte Carlo, PP(70)-GMRES(50) with extra deflation is easily the best approach.  For matrix 1, the time is improved by a factor of about 35 for this best method compared to GMRES(50).  With Matrix 6, one of the two systems does not converge with GMRES(50), thus the dashes, and the other system takes 237 minutes.  This is compared to the average of 5.1 minutes for the best method.
 
\begin{table}
\caption{ Solving linear equations with $24^4$ lattices; matrices of size $n = 3{,}981{,}312$.  Average of two solutions with several approaches. Time is in minutes (per system).}
\begin{center}
\begin{tabular}{|c|c|c||c|c||c|c|c|}  \hline\hline
 & \multicolumn{2}{|c||}{Matrix 1} & \multicolumn{2}{|c||}{Matrix 5} & \multicolumn{2}{|c|}{Matrix 6}   \\ \hline
 Method                 & Time      & MVP's     & Time      & MVP's     & Time      & MVP's             \\  \hline \hline
GMRES(50)               & 215       & 8911      & 113       & 4815      & -         & -                 \\ \hline
PP(50)-GMRES(50)        & 40.4      & 6800      & 22.7      & 3850      & 37.5      & 6325              \\ \hline
Ex defl GMRES(50)       & 27.1      & 1173      & 26.9      & 1140      & 23.7      & 1009              \\ \hline
Defl PP(50)-G(50)       & 11.9      & 1801      & 11.6      & 1976      & 9.5       & 1651              \\ \hline
Ex Defl PP(70)-G(50)    & 6.0       & 1051      & 5.6       & 981       & 5.1       & 911               \\ \hline
\hline
\end{tabular}
\end{center}
\label{Tab:Tr3}
\end{table}
\vspace{.2in}

\subsection{Use of polynomial preconditioning to find eigenvalues}

We have seen that deflating eigenvalues is essential for all methods in this paper, not just for the new method but also for solving the linear equations in standard Hutchison method.  However, the problem of finding the needed small eigenvalues and eigenvectors is difficult and needs to be solved efficiently.  We support this by giving a comparison between standard restarted Arnoldi and the polynomial preconditioned Arnoldi used in this paper.  For the standard Arnoldi, we use a method from~\cite{HRAM} that is mathematically equivalent to implicitly restarted Arnoldi~\cite{So}.  This is run as Arnoldi(120,70) that restarts when the subspace gets to dimension 120 and keeps 70 approximate eigenvectors at each restart.  This is run for 80 cycles which gives about the same number of eigenvalues (about 50) with residual norms below 0.001 as for the PP(70)-Arnoldi method run 150 iterations that was used in Example 3 (there are more very accurate eigenvalues with PP-Arnoldi).  For matrix 1, the result is that restarted Arnoldi takes 21.6 hours and polynomial preconditioned Arnoldi 1.21 hours.  The polynomial preconditioning makes deflation much more practical.

\section{Conclusion}

This paper has a new approach to computing the trace of the inverse of a large QCD matrix.  It proposes a multilevel Monte Carlo method with different degree polynomials.  The polynomials give approximations of different accuracy to the inverse of the matrix.  The lower degree polynomials are from a GMRES iteration and the highest degree polynomial is from polynomial preconditioned GMRES.

This is the first use of polynomials for multilevel Monte Carlo with nonsymmetric matrices.  For this method to be effective, it is essential to deflate eigenvalues from the Monte Carlo steps that have a difference of two polynomials.  This deflation is also new, as is the use of a deflated polynomial to lower the degree of the highest degree polynomial.  Also important to the method is that this highest degree polynomial is a double polynomial with much less orthogonalization expense needed to generate it.  Also implemented is a deflated version of polynomial preconditioned GMRES that is very efficient and is new to QCD calculations.  Computation of eigenvalues and eigenvectors is done with a polynomial preconditioned Arnoldi method that makes deflation much more practical.  

Putting all the pieces mentioned in the previous two paragraphs together gives a multipolynomial Monte Carlo method that is very efficient.  The time for $24^4$ lattices in Example 3 is reduced by more than a factor of six compared to Hutchinson, in spite of the fact that we have also made Hutchinson very efficient. 

We plan future work on this method, including using nonsymmetric Lanczos to develop the polynomials and to compute eigenvalues.  This will likely give a lower degree polynomial than the current double polynomial.  For QCD matrices, this can be implemented efficiently, see~\cite{FroNoeetal95}.

Other future work could be to check the best way to find $Tr(p_3(A))$, with probing as done here or with another Monte Carlo.  It would be interesting to then find what degree is best for that polynomial.  There are other parameters for the algorithm that could be similarly studied, and possibly the choices made more automated.  For example, we should try to figure out how to determine the degree needed for the second polynomial.  It does not need to be as high with the extra deflation.  Also, possibly there should be further experiments with different numbers of polynomials, and a study of whether it is possible for more polynomials to reduce time given the variances and costs.  We also want to try using Z4 noise for the initial vector that generates all polynomials as this would save one linear equations solve.  
Finally, we mention that other quantities besides the trace of the inverse need to be calculated in QCD.  We plan to try our new multi-polynomial approach in different situations.

\vspace{.1in}
\bibliography{morgan}

\begin{thebibliography}{10}

\bibitem{Lan-DR}
A.~M. Abdel-Rehim, R.~B. Morgan, D.~A. Nicely, and W.~Wilcox.
\newblock Deflated and restarted symmetric {L}anczos methods for eigenvalues
  and linear equations with multiple right-hand sides.
\newblock {\em SIAM J. Sci. Comput.}, 32:129--149, 2010.

\bibitem{defbi}
A.~M. Abdel-Rehim, R.~B. Morgan, and W.~Wilcox.
\newblock Deflated {BiCGStab} for linear equations in {QCD} problems.
\newblock {\em Proceedings of Science, LAT2007}, pages 026/1--026/7, 2007.

\bibitem{seed}
A.~M. Abdel-Rehim, R.~B. Morgan, and W.~Wilcox.
\newblock Improved seed methods for symmetric positive definite linear
  equations with multiple right-hand sides.
\newblock {\em Numer. Linear Algebra Appl.}, 21:453--471, 2014.

\bibitem{AsMaOt}
S.~F. Ashby, T.~A. Manteuffel, and J.~S. Otto.
\newblock A comparison of adaptive {C}hebyshev and least squares polynomial
  preconditioning for conjugate gradient methods.
\newblock {\em SIAM J. Sci. Statist. Comput.}, 13:1--29, 1992.

\bibitem{BaBrBrClMa}
R.~Babich, J.~Brannick, R.~Brower, M.~Clark, T.~Manteuffel, S.~McCormick,
  J.~Osborn, and C.~Rebbi.
\newblock Adaptive multigrid algorithm for the lattice {W}ilson-{D}irac
  operator.
\newblock {\em Physical Review Letters}, 105:1--4, 2010.

\bibitem{BaCaGoRe}
J.~Baglama, D.~Calvetti, G.~H. Golub, and L.~Reichel.
\newblock Adaptively preconditioned {GMRES} algorithms.
\newblock {\em SIAM J. Sci. Comput.}, 20:243--269, 1998.

\bibitem{QCDsubtr2}
S.~Baral, R.~B. Morgan, and W.~Wilcox.
\newblock New noise subtraction methods in lattice {QCD}.
\newblock {\em Proceedings of Science, LAT2016}, 2016.

\bibitem{QCDsubtr3}
S.~Baral, T.~Whyte, W.~Wilcox, and R.~B. Morgan.
\newblock Disconnected loop subtraction methods in lattice {QCD}.
\newblock {\em Comput. Phys. Commun.}, 241:64--79, 2019.

\bibitem{BiFr}
S.~Birk and A.~Frommer.
\newblock A deflated conjugate gradient method for multiple right hand sides
  and multiple shifts.
\newblock {\em Numer. Algorithms}, 67:507--529, 2014.

\bibitem{BuEr}
K.~Burrage and J.~Erhel.
\newblock On the performance of various adaptive preconditioned {GMRES}
  strategies.
\newblock {\em Numer. Linear Algebra Appl.}, 5:101--121, 1998.

\bibitem{ChSa}
A.~Chapman and Y.~Saad.
\newblock Deflated and augmented {K}rylov subspace techniques.
\newblock {\em Numer. Linear Algebra Appl.}, 4:43--66, 1997.

\bibitem{Qcdconf2}
D.~Darnell, R.~B. Morgan, and W.~Wilcox.
\newblock Deflation of eigenvalues for iterative methods in lattice {QCD}.
\newblock {\em Nucl. Phys. B (Proc. Suppl.)}, 129:856--858, 2004.

\bibitem{dF}
P.~de~Forcrand.
\newblock Progress on lattice {QCD} algorithms.
\newblock {\em Nucl. Phys. B (Proc. Suppl.)}, 47:228--235, 1996.

\bibitem{PPArn}
M.~Embree, J.~A. Loe, and R.~B. Morgan.
\newblock Polynomial preconditioned {A}rnoldi with stability control.
\newblock {\em SIAM J. Sci. Comput.}, 43:A1--A25, 2021.

\bibitem{ErBuPo}
J.~Erhel, K.~Burrage, and B.~Pohl.
\newblock Restarted {GMRES} preconditioned by deflation.
\newblock {\em J. Comput. Appl. Math.}, 69:303--318, 1996.

\bibitem{FrVu}
J.~Frank and C.~Vuik.
\newblock On the construction of deflation-based preconditioners.
\newblock {\em SIAM J. Sci. Statist. Comput.}, 23:442--462, 2001.

\bibitem{FrKaKrLeRo}
A.~Frommer, K.~Kahl, S.~Krieg, B.~Leder, and M.~Rottman.
\newblock Adaptive aggregation based domain decomposition multigrid for the
  lattice {W}ilson {D}irac operator.
\newblock {\em SIAM J. Sci. Comput.}, 36:A1581--A1608, 2014.

\bibitem{FrKhRH}
A.~Frommer, M.~N. Khalil, and G.~Ramirez-Hidalgo.
\newblock A multilevel approach to variance reduction in the stochastic
  estimation of the trace of a matrix.
\newblock {\em SIAM J. Sci. Comput.}, 44:A2536--A2556, 2022.

\bibitem{FroNoeetal95}
Andreas Frommer, Bertold N\"{o}ckel, Stephan G\"{u}sken, Thomas Lippert, and
  Klaus Schilling.
\newblock {Many masses on one stroke: Economic computation of quark
  propagators}.
\newblock {\em Int. J. Mod. Phys.}, C6:627--638, 1995.

\bibitem{GaStOr}
A.~S. Gambhir, A.~Stathopoulos, and K.~Orginos.
\newblock Deflation as a method of variance reduction for estimating the trace
  of a matrix inverse.
\newblock {\em SIAM J. Sci. Comput.}, 39:A532–A558, 2017.

\bibitem{Gi08}
M.~B. Giles.
\newblock Multilevel {M}onte {C}arlo path simulation.
\newblock {\em Oper. Res.}, 56:607--617, 2008.

\bibitem{Gi15}
M.~B. Giles.
\newblock Multilevel {M}onte {C}arlo methods.
\newblock {\em Acta Numer.}, 24:259--328, 2015.

\bibitem{GiHaNaSc}
L.~Giusti, T.~Harris, A.~Nada, and S.~Schaefer.
\newblock Frequency-splitting estimators of single-propagator traces.
\newblock {\em Eur. Phys. J. C}, 79:586 (17 pages), 2019.

\bibitem{QCDsubtr}
V.~Guerrero, R.~B. Morgan, and W.~Wilcox.
\newblock Eigenspectrum noise subtraction methods in lattice {QCD}.
\newblock {\em Proceedings of Science, LAT2009}, pages 041/1--041/8, 2009.

\bibitem{HaTr}
E.~Hallman and D.~Troester.
\newblock A multilevel approach to stochastic trace estimation.
\newblock {\em Linear Algebra Appl.}, 638:125--149, 2022.

\bibitem{He98}
S.~Heinrich.
\newblock Monte {C}arlo complexity of global solution of integral equations.
\newblock {\em J. Complexity}, 14:151--175, 1998.

\bibitem{HutchTrace}
M.~F. Hutchinson.
\newblock A stochastic estimator of the trace of the influence matrix for
  {L}aplacian smoothing splines.
\newblock {\em Commun. Stat. Simul. Comput.}, 18:1059--1076, 1989.

\bibitem{KhYe}
S.~A. Kharchenko and A.~Y. Yeremin.
\newblock Eigenvalue translation based preconditioners for the {GMRES}(k)
  method.
\newblock {\em Numer. Linear Algebra Appl.}, 2:51--77, 1995.

\bibitem{LaSt}
J.~Laeuchli and A.~Stathopoulos.
\newblock Extending hierarchical probing for computing the trace of matrix
  inverses.
\newblock {\em SIAM J. Sci. Comput.}, 42:A1459--A1485, 2020.

\bibitem{PolyQCD}
P.~Lashomb, R.~B. Morgan, T.~Whyte, and W.~Wilcox.
\newblock High-degree polynomial noise subtraction for disconnected loops.
\newblock ArXiv:2306.06188, 2023.

\bibitem{LiXiVeYaSa}
R.~Li, Y.~Xi, E.~Vecharynski, C.~Yang, and Y.~Saad.
\newblock A thick-restart {L}anczos algorithm with polynomial filtering for
  {H}ermitian eigenvalue problems.
\newblock {\em SIAM J. Sci. Comput.}, 38:A2512--A2534, 2016.

\bibitem{PPG}
Q.~Liu, R.~B. Morgan, and W.~Wilcox.
\newblock Polynomial preconditioned {GMRES} and {GMRES-DR}.
\newblock {\em SIAM J. Sci. Comput.}, 37:S407--S428, 2015.

\bibitem{PPGStable}
J.~A. Loe and R.~B. Morgan.
\newblock Toward efficient polynomial preconditioning for {GMRES}.
\newblock {\em Numer. Linear Algebra Appl.}, 29:1--21, 2021.

\bibitem{Lusch}
M.~L\"uscher.
\newblock Local coherence and deflation of the low quark modes in lattice
  {QCD}.
\newblock {\em JHEP}, 0707:081, 2007.

\bibitem{GMRES-E}
R.~B. Morgan.
\newblock A restarted {GMRES} method augmented with eigenvectors.
\newblock {\em SIAM J. Matrix Anal. Appl.}, 16:1154--1171, 1995.

\bibitem{GMRES-DR}
R.~B. Morgan.
\newblock {GMRES} with deflated restarting.
\newblock {\em SIAM J. Sci. Comput.}, 24:20--37, 2002.

\bibitem{Qcdconf}
R.~B. Morgan and W.~Wilcox.
\newblock Deflation of eigenvalues for {GMRES} in lattice {QCD}.
\newblock {\em Nucl. Phys. B (Proc. Suppl.)}, 106:1067--1069, 2002.

\bibitem{gproj}
R.~B. Morgan and W.~Wilcox.
\newblock Deflated iterative methods for linear equations with multiple
  right-hand sides.
\newblock arXiv:math-ph/0405053v2, 2004.

\bibitem{HRAM}
R.~B. Morgan and M.~Zeng.
\newblock A harmonic restarted {A}rnoldi algorithm for calculating eigenvalues
  and determining multiplicity.
\newblock {\em Linear Algebra Appl.}, 415:96--113, 2006.

\bibitem{Ni}
R.~A. Nicolaides.
\newblock Deflation of conjugate gradients with applications to boundary value
  problems.
\newblock {\em SIAM J. Numer. Anal.}, 24:355--365, 1987.

\bibitem{PadeStMaJoMa}
M.~L. Parks, E.~de~Sturler, G.~Mackey, D.~D. Johnson, and S.~Maiti.
\newblock Recycling {K}rylov subspaces for sequences of linear systems.
\newblock {\em SIAM J. Sci. Comput.}, 28:1651--1674, 2006.

\bibitem{RoStOr}
E.~Romero, A.~Stathopoulos, and K.~Orginos.
\newblock Multigrid deflation for lattice qcd.
\newblock {\em J. Comput. Phys.}, 409:109356, 2020.

\bibitem{Sa84b}
Y.~Saad.
\newblock {C}hebychev acceleration techniques for solving large nonsymmetric
  eigenvalue problems.
\newblock {\em Math. Comp.}, 42:567--588, 1984.

\bibitem{Sa87b}
Y.~Saad.
\newblock Least squares polynomials in the complex plane and their use for
  solving sparse nonsymmetric linear systems.
\newblock {\em SIAM J. Numer. Anal.}, 24:155--169, 1987.

\bibitem{Sa95B}
Y.~Saad.
\newblock Analysis of augmented {K}rylov subspace techniques.
\newblock {\em SIAM J. Matrix Anal. Appl.}, 18:435--449, 1997.

\bibitem{So}
D.~C. Sorensen.
\newblock Implicit application of polynomial filters in a $k$-step {A}rnoldi
  method.
\newblock {\em SIAM J. Matrix Anal. Appl.}, 13:357--385, 1992.

\bibitem{LaStOr}
A.~Stathopoulos, J.~Laeuchli, and K.~Orginos.
\newblock Hierarchical probing for estimating the trace of the matrix inverse
  on toroidal lattices.
\newblock {\em SIAM J. Sci. Comput.}, 35:S299--S322, 2013.

\bibitem{StOr}
A.~Stathopoulos and K.~Orginos.
\newblock Computing and deflating eigenvalues while solving multiple right hand
  side linear systems in quantum chromodynamics.
\newblock {\em SIAM J. Sci. Comput.}, 32:439--462, 2010.

\bibitem{ThDoLiYe}
C.~Thron, S.~Dong, K.~F. Liu, and H.~P. Yeng.
\newblock Padé-{Z}2 estimator of determinants.
\newblock {\em Phys. Rev. D}, 57:1642--1653, 1998.

\bibitem{WhStRoOr}
T.~Whyte, A.~Stathopoulos, E.~Romero, and K.~Orginos.
\newblock Optimizing shift selection in {M}ultilevel {M}onte {C}arlo for
  disconnected diagrams in {L}attice {QCD}.
\newblock ArXiv:2212.04430, 2022.

\bibitem{Walt6}
W.~Wilcox.
\newblock Perturbative subtraction methods.
\newblock {\em Nucl. Phys. B (Proc. Suppl.)}, 83-84:834--836, 2000.

\end{thebibliography}

\end{document}